# Measuring general relativity effects in a terrestrial lab by means of laser gyroscopes

**N Beverini[1,2], M Allegrini[1,2], A Beghi[3], J Belfi[1], B. Bouhadef[2], M Calamai[1,2], G Carelli[1,2], D Cuccato[3], A Di Virgilio[2], E Maccioni[1,2], A Ortolan[4], A Porzio[5,6], R Santagata[2,7], S Solimeno[6,8] and A Tartaglia[9]**

[1]Dipartimento di Fisica, Università di Pisa and CNISM, Pisa, Italy;
[2]INFN, Sezione di Pisa, Pisa, Italy;
[3]Department of Information Engineering, Università di Padova, Padua, Italy
[4]INFN,Laboratori di Legnaro, Legnaro (PD), Italy;
[5]CNR-SPIN, Naples, Italy;
[6]INFN, Sezione di Napoli, Naples, Italy;
[7]Dipartimento di Fisica, Università di Siena, Siena, Italy;
[8]Dipartimento di Fisica, Università di Napoli "Federico II", Naples, Italy
[9]Politecnico of Torino and INFN, Turin, Italy

beverini@df.unipi.it

**Abstract.** GINGER is a proposed tridimensional array of laser gyroscopes with the aim of measuring the Lense-Thirring effect, predicted by the General Relativity theory, in a terrestrial laboratory environment. We discuss the required accuracy, the methods to achieve it, and the preliminary experimental work in this direction.

## 1. Introduction

Large frame optical laser gyroscopes (*gyrolasers*) are presently the most accurate instruments for measuring rotation rates with respect to an inertial reference systems [1]. They exploit the Sagnac effect: if a ring laser, operating on a single mode, rotates with respect to an inertial reference system, the radiation beam traveling in the rotation direction sees a round-trip path longer than the counter-propagating one. As a consequence, the stationary frequencies of the radiation emitted in the two directions differ by:

$$\left| f_+ - f_- \right| = \frac{4}{\lambda p} \mathbf{A} \cdot \mathbf{\Omega} \qquad (1)$$

where **A** is the area vector corresponding to the surface encircled by the beams, $p$ is the perimeter of the geometrical path followed by the light rays, **Ω** is the effective angular velocity vector with respect to the "fixed stars", and λ is the laser radiation wavelength. This result, which is easily demonstrated in classical radiation theory, can also be confirmed in a complete general relativistic treatment. The frequency difference, called the *Sagnac frequency*, can be measured by observing on a photodiode the beat note between the two counter-rotating beams.

The resolution of a gyrolaser is ultimately limited by quantum noise. The uncertainty on the angular frequency produced by quantum noise can be evaluated as:

$$\Omega_{sn} = \frac{c\,p}{2AQ}\sqrt{\frac{h\nu_L}{P_{out}T}} \qquad (2)$$

where $c$ is the speed of light, $Q$ is the quality factor of the optical cavity, $P_{out}/(h\nu_L)$ is the number of photons detected per unit time, and $T$ is the observation time. In a ring cavity, where the losses are essentially related to the mirror reflectivity, the $Q$ factor of the cavity is proportional to its length. Thus, the quantum noise decreases as the square of the linear dimension of the ring, and a larger ring cavity should have a better resolution.

The best gyrolaser operating at the moment is the Großring G, which is located in the Bayern geodetic observatory in Wettzell. It is a square ring laser with 4-meter side, mounted on a monolithic structure made of Zerodur (a glass with practically null expansion coefficient at room temperature), equipped with supermirrors with a reflectivity larger than 99.9992% ($Q = 3.5 \cdot 10^{12}$). G has achieved a resolution near to the quantum noise limit, in about 1 hour of measurement time, reaching an optimal sensitivity level of $3 \cdot 10^{-13}$ rad/s. For longer times random walk noise dominates, but the stability is anyway good enough to allow for a careful estimation of the long term fluctuation of the orientation of the rotation axis of the Earth (the Chandler wobble) [2].

Larger laser gyros (UG1 and UG2) have been built and tested at the Canterbury University in New Zealand [3]. They have a heterolithic structure and suffer from Earth's crust strains, thermo-elastic deformations and atmospheric pressure variations, so much that the other sources of noise largely overcome the quantum noise. Their final resolution turns out to be orders of magnitude worse of G, in spite of the much larger area (respectively 367 and 834 m$^2$).

In this paper we will present the work being done in order to build a tridimensional array of laser gyroscopes that can improve over the present G resolution in order to enable the measurement of the inertial reference frames dragging induced by the angular momentum of the rotating Earth. This drag (named after Lense and Thirring) is a general relativistic effect. In terms of equivalent rotation it is in the order of a few $10^{-14}$ rad/s. In §2, we will present shortly the theoretical frame in which this effect appears. The requirement on the gyro system will be presented and the way to achieve the desired results will be analysed in §3. Finally in §4 we will describe the present research and development activity in order to demonstrate the feasibility of the experiment.

**2. General Relativity effects induced on an Earth based gyroscope**

Applying the formalism of Einstein general relativity, the solutions of general relativistic field equations show that the gravitational field of a rotating source, such as the Earth or the Sun, is made, to lowest order, of two components: a gravito-electric field, due to the mass, which, at the lowest approximation order, coincides with the Newtonian field, and a gravito-magnetic field, due to the mass currents. In the case of a rotating source the gravito-magnetic contribution cannot be eliminated by a global coordinate transformation and has a dipolar structure; its amplitude is proportional to the angular momentum of the source. The expected rotation signal seen by a laser gyroscope located in a laboratory on the Earth surface with co-latitude $\theta$ and with the axis contained in the meridian plane at an angle $\psi$ with respect to the zenith, is given by:

$$\nu \cong 4\frac{S}{\lambda p}\Omega_T\left[\cos(\theta+\psi) - 2\frac{GM_T}{c^2 R_T}\sin\theta\sin\psi \right.$$
$$\left. + \frac{GI_T}{c^2 R_T^3}(2\cos\theta\cos\psi + \sin\theta\sin\psi)\right] \qquad (3)$$

where $G$ is Newton's gravitational constant, $\Omega_T = 7.29 \cdot 10^{-5}$ rad/s is the rotation speed of the Earth, $M_T$ the mass of the Earth, $R_T$ the Earth mean radius, and $I_T \approx 2/5\ M_T\ R_T^2$ the Earth moment of inertia.

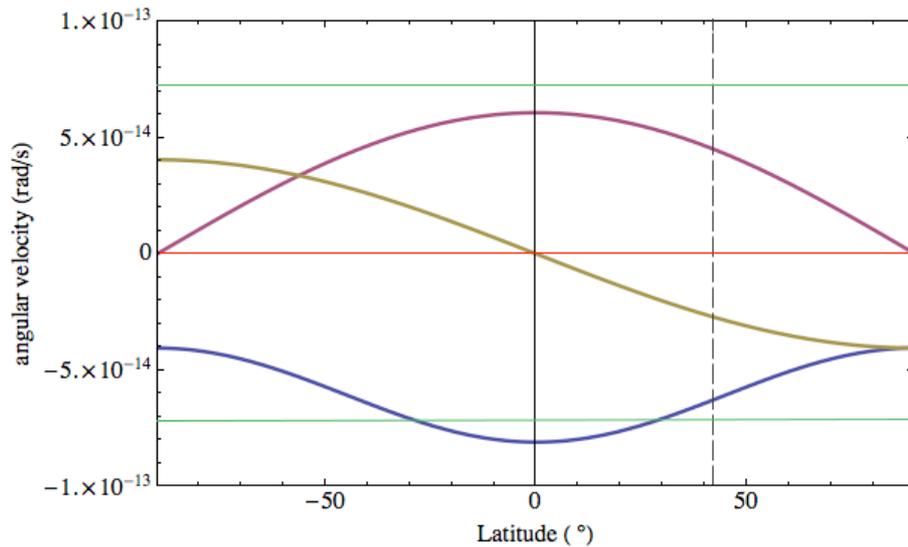

Fig.1 - The amplitude of the general-relativistic signals (de Sitter and LT precessions) on the surface of the Earth, as a function of the latitude. The upper line (green *on line*), the second line (purple *on line*), and the lower line (blue *on line*) correspond to a gyro axis orientation in the meridian plane, respectively in local radial (zenithal) direction, in the direction of Earth's rotation axis, in the local North-South direction. The light-green horizontal lines indicate an equivalent angular speed equal to $\pm 10^{-9}$ times the Earth rotation rate. The vertical dashed line indicates the Gran Sasso Laboratories latitude.

Developing the square bracket we obtain three frequencies. The first corresponds to the classical Sagnac signal expressing the purely kinematical effect of the rotation of the gyro (which coincides with the rotation of the whole Earth); the second beat frequency is produced by the coupling of the gravito-electric field with the rotation of the laboratory (carried by the rotating Earth (*geodetic* or *De Sitter precession*); the third contribution is produced by the gravito-magnetic field due to the angular momentum of the Earth (*Lense-Thirring precession*, LT). Both general-relativistic effects are depressed with respect to the Sagnac effect by a factor of the order of magnitude of the ratio between $R_T$ and the Schwarzschild radius of the Earth. The amplitude of the relativistic precessions as a function of the laboratory latitude is shown in Fig.1 for different orientations of the gyro's axis. We note that along the radial direction (green curve on the graph) the de Sitter precession is null and the signal only contains the gravito-magnetic term.

At the moment, the general relativistic effects have been tested by space experiments only. The Gravity-Probe B (GP-B) experiment was firstly proposed in 1970. After a long gestation the experiment started in 2004. It consisted of a terrestrial satellite in polar orbit, carrying on board four superconducting gyroscopes and a telescope. The data taking lasted one year and a half; the final data analysis was published in 2011 [4]. It reports an accuracy of 0.28% on the geodetic (de Sitter) precession, but only 19% on LT. Serious problems arose for unexpected and non-reducible systematic inconveniences in one of the superconducting gyroscopes.

Another experiment in space did initially not consist in a dedicated mission, but raher in an extremely accurate analysis, made by Ciufolini and Pavlis, of the data gathered by the laser ranging of the LAGEOS and LAGEOS II satellites. These two satellites, launched at different epochs, are located in almost identical orbits with different inclinations: 52.6° and 109.8°. Measuring the precession of the orbital nodes, and averaging over some years, the above authors found an evidence of the LT effect within 10% accuracy [5]. A new dedicated satellite (LARES) was launched on February 13[th], 2012, and is now taking data. It consists of a massive very high density tungsten alloy sphere, 36 cm diameter, carrying 92 cat-eye reflectors [6]. The goal of the experiment is to reach a 1-2% accuracy in the measurement of the LT precession.

A measurement of the LT precession in a ground laboratory using laser gyroscopes can be considered as complementary to the space based ones [7]. The setup and principle of the experiment is indeed entirely different. In space the observer is in geodetic motion (free fall), while in a terrestrial laboratory the observer is in a non-inertial (accelerated) motion. Moreover in the ground experiment the effective precession is measured locally, while in space the effect is integrated along the whole orbital motion, and requires a very accurate knowledge of the gravitational field along the orbits. Furthermore an apparatus in a terrestrial laboratory is much more accessible for modifying the measurement parameters. Last but not least, the cost of an Earth-based experiment is much lower than for a space-based one.

As a fact, we expect a positive synergy from the collaboration and the exchange of data between the two approaches. The complementary experimental methods make it more likely the capability to highlight deviations from the results anticipated by GR, if they exists. Deviations are predicted by various alternative theories to GR, such as string theories and others, developed mainly to justify the 73% of the energy content of the universe of which the origin is not clear (besides the other 23% of yet unidentified dark matter).

## 3. The gyroscope apparatus for LT precession measurement

*3,1 Signal accuracy*

From the results presented in the previous section we have found that relativistic precessions will give a correction of the order of 1 part in $10^9$ on the Sagnac frequency measured by a gyroscope. Then, we need to achieve a resolution of our apparatus of the order of $10^{-10}$, which means an improvement of a bit more than one order of magnitude with respect to the present G-ring performances. By extrapolating from G-ring performances, using mirrors of comparable quality and increasing the gyro cavity side length to 6 – 8 m should improve the quantum noise limit by a factor 3 - 4. Another factor 3 – 4 might be gained by improving the long-term stability up to 1 day.

In any case, however, it would not be possible to build a "big G" on a similar monolithic structure. The special glass with zero expansion coefficient would have prohibitive cost and, moreover, is impossible to use it for building a monolithic platform in such a large scale. It is then necessary to replace the lost mechanical rigidity with sophisticated diagnostic systems and active stabilization of the mirror positions. Area and perimeter of the optical cavity, and the emission wavelength must be stabilised with accuracy better than $10^{-10}$. Locating the gyro in an underground laboratory can enhance the long-term stability in order to reduce the low-frequency noise induced near the surface by seism and by meteorological events. It is also compulsory to subtract from the detected signal all the known geophysical perturbations (Earth tides, ocean tides, polar motion, tectonic motions, etc.).

Eq. (1) expresses the idealized relation between the measured Sagnac frequency difference and the ring cavity rotation rate. In the real situation, however, we must consider the presence of perturbing effects, which can be resumed by the following equation:

$$\Delta f = k_s (1 + k_A) \Omega + \Delta f_0 + \Delta f_{BS} \qquad (4)$$

where $k_s = 4A/(\lambda p)$ is the geometrical form factor of Eq.(1), $k_A$ takes into account the changes in the optical length of the cavity due to the dispersion properties of the plasma discharge, $\Delta f_0$ is the null shift, due to the non-reciprocity of the optical cavity, while $\Delta f_{BS}$ takes into account the frequency pulling due to the coupling between the two beams traveling in opposite directions. This coupling is actually given by radiation backscattering on the mirrors. To gain the necessary precision, all these effects must be identified and carefully controlled.

*3.2 Optimization of the ring geometry*

To keep any change in the form factor below $10^{-10}$ appears very hard job to be performed. The sides of the cavity ring and the relative angles must be controlled with a very high accuracy. This task can be anyhow afforded by adopting suitable cavity geometry.

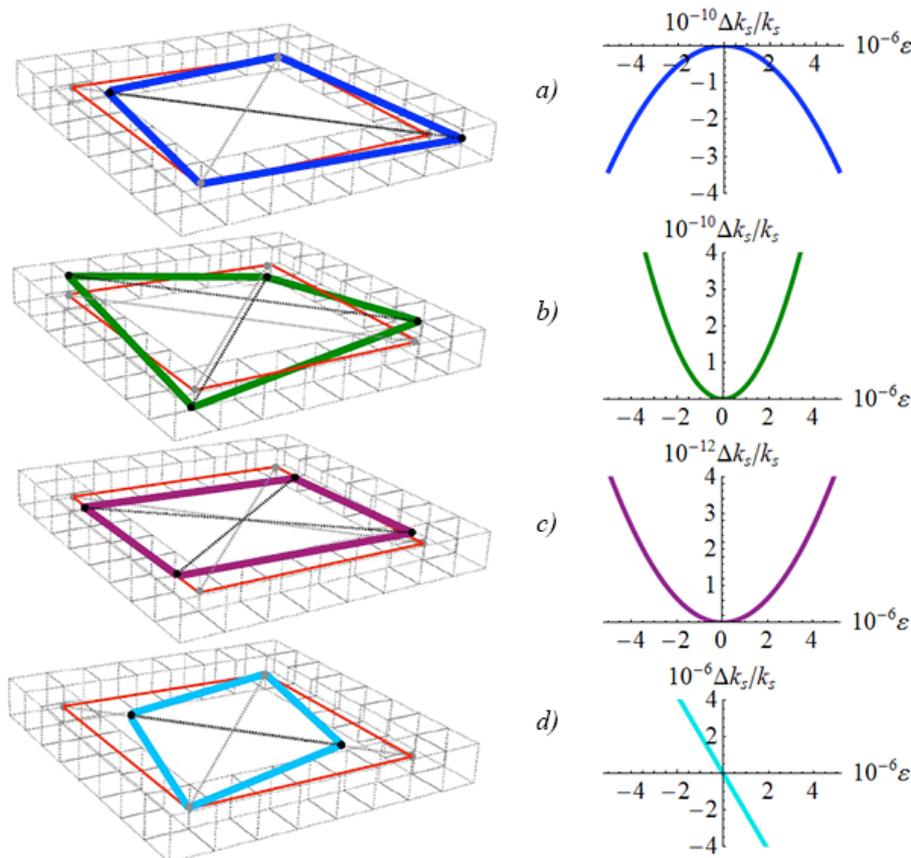

Fig. 2 –Left: Visualization of the deformation degrees of freedom of a square cavity. *a)* rigid displacement in the ring plane of the two diagonals, one with respect to the other, keeping their relative angle fixed (2 d.o.f.); *b)* as the previous, but moving in the orthogonal plane; *c)* relative rotation of the two diagonals; *d)* Stretching the diagonal lengths (2 d.o.f.).

Right: Fractional changes in the geometrical form factor $k_s = 4A/(\lambda p)$ as a function of the deviation from exact geometry for the correspondent degree of freedom.

Let us consider a square ring cavity delimited by four identical spherical mirrors. Thanks to mirror spherical symmetry, the radiation path inside the cavity is univocally defined by the positions of the four centres of curvature (CC). Then, the whole optical cavity, which we suppose having a shape very near to a plane square, can be defined by 12 degrees of freedom (d.o.f.), which are linear combinations of linear CC displacements (or equivalently of mirror displacements). It is possible to define these d.o.f. so that six of them correspond to cavity rigid body movements that do not affect the form factor. We choose the other six d.o.f. so that two will correspond to stretching movements of the ring diagonals, three to relative rigid movements of the two diagonals one with respect to the other (two in the ring plane and one in the orthogonal plane) and one producing a relative rotation of the diagonals (Fig. 2). It is easy to prove that, around the "perfect square" position, the form factor is linearly sensitive to the stretching d.o.f. , but has only a quadratic sensitivity to the other d.o.f. .

The diagonal length can be quite easily controlled at high accuracy, because the two couples of opposite mirrors define two Fabry-Pérot cavities whose length can be locked to the wavelength of an injected laser beam frequency stabilised to an atomic reference. It is possible to prove that, freezing these two d.o.f's, the *A/p* ratio becomes a quadratic form in the deviations of the other d.o.f's from the "perfect square" geometry. Then, an accuracy of the order of $10^{-5}$ on the control of these other d.o.f. is sufficient to obtain $10^{-10}$ accuracy of the *A/p* ratio, provided that the accuracy in the two diagonal lengths is better than $10^{-10}$.

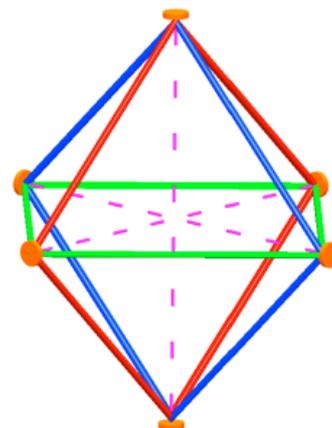

Fig. 3 - The octahedral multi-ring gyro configuration.

The observed Sagnac frequency is actually sensitive also to the angle between **A** and **Ω** i.e. to the absolute orientation of the gyro with respect the "fixed stars". To achieve the required accuracy this angle should be known with an accuracy much better than 1 nrad: clearly an impossible task in an underground laboratory. This problem can be overcome by using a set of three (or more) laser gyros with different axial orientation in order to reconstruct the modulus of the angular rotation rate that is invariant with respect the absolute orientation. In this way it is always necessary to measure with very high precision the relative dihedral angle between the planes of the rings. A further help in the multi-ring configuration might come assembling the rings in a highly symmetric configuration. Let us consider an octahedral structure in which six mirrors define three reciprocally orthogonal ring laser cavities, nested with each mirror shared between two rings. In this way, each diagonal of the octahedron is shared by two different cavities. Locking the three diagonals of the octahedron to a high-accuracy reference laser at an identical value and optimizing the three rings shape will also give strong constrains to the reciprocal orientations, relaxing the required a priori accuracy in the dihedral angles.

*3.3 The backscattering*

For small rotation rate, the difference between the resonance frequency of the two resonant waves traveling in opposite directions in the optical cavity is smaller than the resonance width and the radiation back-reflected from the mirrors then can be amplified by the active medium. This corresponds to the classical problem of two pendulums of similar frequency coupled together. The two proper resonance frequencies are shifted and, beyond a threshold, the two oscillators are locked. This is a well-known effect that limits the performance of the commercial laser gyroscopes. To achieve enough sensitivity for using the Earth rotation rate as a bias it was necessary to build large frame (>1 $m^2$) rings and to use "supermirrors" with a very high reflectivity (>99,999%). The effect of back-scattering scales quickly with the ring size: the longer is the ring side the larger is the frequency bias due to the Earth; the higher is the cavity finesse, the smaller is the solid angle acceptance of the back scattered radiation. Finally, a scale factor proportional to the $4^{th}$ power of the linear dimension was estimated [7]. We remark that the coupling factor is produced by the interference between the waves back-scattered by each single mirror; its effective value is then very sensitive to the relative positions of the mirrors.

Besides producing a frequency pulling $\Delta f_{BS}$, backscattering can be responsible also for a null-shift. Actually, it is a dissipative phenomenon, and can then introduce non-reciprocal effects in the laser operation. As a fact, experimentally it has been observed that the intensity of the two traveling beams is usually quite different, even more than 10% with the smaller gyroscope (between 1 and 2 m side) [1], but still of the order of 1% with the G-ring (4 m side).

All these considerations impose to make our set of ring gyroscopes as large as possible, compatibly with the space available in an underground laboratory. As we will discuss in more details in the next section, making use of an accurate diagnostic of the laser parameters, the application of a Kalman filter will allow to improve long term stability.
.
**4. The approach to GINGER experiment**

*4.1 Laser stability control*
In the last years in the Pisa laboratories we have worked with the G-Pisa gyro [9,10], a square ring with 1.35 m side, initially located at the Virgo site and later in S.Piero INFN laboratory.

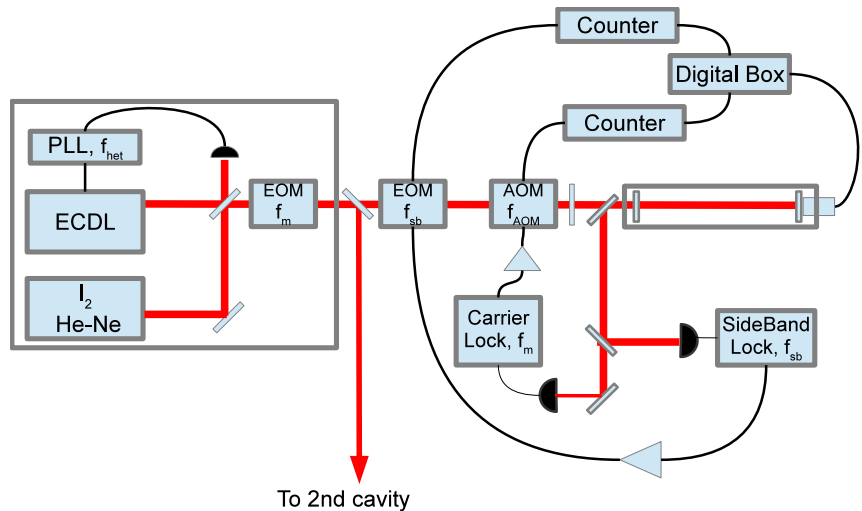

Fig.4 – Scheme of the Pound-Drever-Hall double modulation circuit for the absolute measurement and the stabilisation of the two diagonal optical cavities.

Exploiting the acquired expertise, we have now designed a new structure, called GP2 that will be devoted specifically to the optimization and the stabilization of the gyro's form factor. GP2 will be placed in the INFN Pisa laboratory and will have a square ring cavity with 1.60 m side. The axis of the ring will be parallel to the Earth rotation axis, in order to maximize the Sagnac frequency and minimize the effect of ground tilting which, in that configuration is quadratic in the angle amplitude. The mirrors are fixed in four towers mounted at the corners of a granite slab weighing 1100 kg and are linked together by steel vacuum pipes both along the sides and along the diagonals of the square. Three towers are equipped with a PZT linear actuator working along the diagonal direction, while the other tower is equipped with a 3-dimensional actuator. It is then possible to act upon all 6 deformation d.o.f.'s .

The diagonal lengths will be measured with respect to an interrogating high-stability laser by a two frequency Pound-Drever-Hall circuit [11], in order to lock the cavities to the reference at a same absolute value (Fig.4). The relative phase will be locked with the first frequency modulation circuit, while the order of interference will be measured by a second modulation circuit operating at the frequency of the cavity free spectral range (or of a harmonic of it).

On this apparatus we will test also the technique for reducing the effect of back-scattering. We have built [12] a model of the gas laser operation that is based on the Lamb formalism [13] extended by Aronowitz [14,15] to ring cavities. In this theory, the laser action has been modelled at the $3^{rd}$ order in atomic polarization, producing a coupled set of equations relating the field amplitude and phase of the two opposite laser beams in terms of a set of effective parameters. By identifying these parameters in terms of observables that can be experimentally measured, we have built a Kalman filter which can be applied to deconvolute the true Sagnac frequency from the recorded data. The effectiveness of this approach has been already proved by analysing the data taken on G-Pisa. The preliminary results are quite straightforward (Fig. 5). We have operated on a 2-day recording time, estimating the instantaneous Earth rotation velocity in 1-second samples, both through the standard frequency reconstruction of the raw data and through the application of our Kalman filter. The filtered data show a reduction of one order of magnitude of the standard deviation with respect to the raw data. Moreover, while the raw data analysis indicates the presence of an important null-shift coefficient, the latter is almost totally removed by the application of the Kalman filter.

*4.2 Gran Sasso site analysis*
The GINGER experiment requires an underground laboratory large enough to host the apparatus. The Gran Sasso laboratories satisfy this criterion. The laboratories lay aside the L'Aquila – Teramo highway tunnel through the Gran Sasso massif, more than 1000 m deep under the surface. It harbors three very large experimental halls (some hundreds meters length and 20x20 $m^2$ section) and many smaller connecting tunnels. The final location of our apparatus is not yet defined, but hall B seems to be convenient.

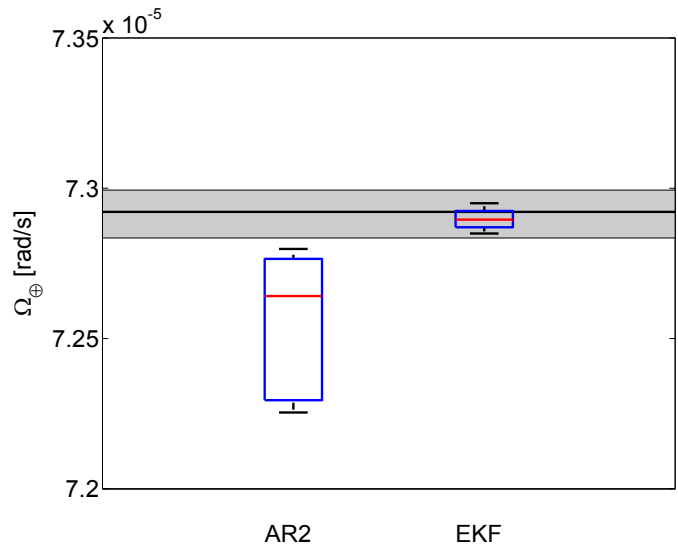

Fig. 5 – Comparison between the distributions of Earth angular velocity appraisals in two days of recording, by standard frequency evaluation of raw data (AR2) and by the Extended Kalman Filter estimate (EKF). The rotation velocity was estimated on 1-second samples. The boxes (blue on line) correspond to two-variance intervals, the bar to three-variance interval; the red line inside the box corresponds to the mean value. The black horizontal line is the Earth angular velocity and the grey box represents the scale factor uncertainties of G-PISA, due to geometric and orientation tolerances.

The site must be carefully analysed in order to characterize the local seismic noise, in particular at the very low frequencies, around 1 mHz and below, which are outside the operative range of the seismometers. For this purpose we have recently brought there the G-Pisa gyro and produced a sample of the local noise level. The instrument will now be remounted in hall B on an enlarged frame (around 3.50 m side). This new "G-Pisa-Plus" (GP+) will have enough sensitivity to make a definitive validation of the site and to perform rotational measurements of interest for geophysics and geodesy.

**5. Conclusions**

The Earth rotation rate measured by the laser gyroscope in a ground laboratory will be compared with the Earth rotation rate measured in the cosmic inertial frame by Very Long Baseline Interferometry, whose relative accuracy is in the range of a few $10^{-11}$, produced by I.E.R.S (International Earth Rotation Service). The geodetic and LT precessions will be given by the difference between the two data.

The analysis that we have presented here suggests that it will be possible to achieve an accuracy of $10^{-10}$, and may be better, by a multi-ring laser gyroscope apparatus with linear size of the order of 8-10 m, placed in an underground laboratory. An intense theoretical and experimental work is anyhow necessary.